\documentclass[pre,twocolumn,preprintnumbers,amsmath,amssymb]{revtex4}

\usepackage{graphicx}
\usepackage{dcolumn}
\usepackage{bm}
\usepackage{parskip}
\usepackage{color}
\usepackage{subfigure}
\usepackage{amssymb}
\usepackage{amsmath}
\usepackage{amssymb}
\usepackage{graphicx}

\def\be{\begin{eqnarray}}
\def\ee{\end{eqnarray}}
\def\be{\begin{equation}}
\def\ee{\end{equation}}
\begin{document}

\title {Exact Critical Casimir Amplitude of Anisotropic Systems from Conformal Field Theory and Self-Similarity of Finite-Size Scaling Functions in {\bf $d\geq 2$} Dimensions}
\author{Volker Dohm and Stefan Wessel}

\affiliation{Institute for Theoretical Physics, RWTH Aachen University, 52056 Aachen, Germany}

\date {January 08, 2021}

\begin{abstract}
The exact critical Casimir amplitude is derived for anisotropic systems within the $d=2$ Ising universality class by combining conformal field theory (CFT) with anisotropic $\varphi^4$ theory. Explicit results are presented for the general anisotropic scalar $\varphi^4$ model and for the fully
anisotropic triangular-lattice Ising model in finite rectangular and infinite strip geometries with periodic boundary conditions (PBC). These results demonstrate the validity of multiparameter universality for confined anisotropic systems and the nonuniversality of the critical Casimir amplitude. We find an unexpected complex form of self-similarity of the anisotropy effects near the instability where weak anisotropy breaks down. This can be traced back to the property of modular invariance of isotropic CFT for $d=2$. More generally, for $d>2$ we predict the existence of self-similar structures of the finite-size scaling functions of $O(n)$-symmetric  systems with planar anisotropies and  PBC both in the critical region for $ n \geq 1$ as well as in the Goldstone-dominated low-temperature region for $ n \geq 2$.
%
\end{abstract}
\maketitle
Fluctuation-induced thermodynamic forces are ubiquitous in confined condensed matter systems~\cite{kardar}. They exist in both isotropic systems such as fluids, superfluids, and binary liquid mixtures~\cite{krech,gambassi} as well as in anisotropic systems such as liquid crystals~\cite{kardar,ajdari1991}, superconductors~\cite{wil-1}, and compressible solids~\cite{dohm2011}. Near a critical point, so-called critical Casimir forces~\cite{krech,gambassi} arise from long-range critical fluctuations, which generate a universal finite-size critical behavior that can be classified in universality classes with universal critical exponents \cite{priv}.
Within a universality class there exist subclasses \cite{dohm2008,dohm2018} of isotropic and weakly anisotropic $d$-dimensional systems -- the latter have $d$ independent nonuniversal correlation-length amplitudes in $d$  principal directions.
While the Casimir force amplitude at criticality is widely believed to be a universal quantity
\cite{affleck,bloete,priv,pri,krech,gambassi,dubail,privman1990},
this is not valid for weakly anisotropic $O(n)$-symmetric systems with an $n$-component order parameter
in $2<d<4$ dimensions
\cite{cd2004,dohm2006,dohm2008,kastening-dohm,dohm2018}.
Furthermore, low-temperature Casimir forces
due to
Goldstone modes \cite{dohm2013}
exhibit nonuniversal anisotropy effects \cite{dohm2018}.
Recently the hypothesis of multiparameter universality for weakly anisotropic systems has  been put forward \cite{dohm2018} but no proof has been given for confined systems and no detailed analysis has been performed near the instability where weak anisotropy breaks down.
In particular,  the universality properties of the critical Casimir amplitude of finite  anisotropic systems in
$d=2$
have remained unexplored in the literature.

Two-dimensional systems are of fundamental theoretical  interest since conformal field theory (CFT) is capable of deriving rigorous results for critical Casimir amplitudes of {\it isotropic} systems on a strip
\cite{affleck,cardy1987,dubail,bloete,cardy1986} and for the partition function at the critical temperature $T_c$ on a parallelogram
\cite{cardy1986-270,franc1987,franc1997,itz,cardyconform}.
In this Letter our focus is on the critical Casimir force in
weakly anisotropic ($d=2,n=1$) Ising-like systems for which CFT has not made any prediction so far.
We show how to combine an exact result of CFT for the isotropic Ising model on a torus \cite{franc1987,franc1997} with an exact
shear transformation of
anisotropic $\varphi^4$
theory \cite{dohm2006} which, on the basis of multiparameter universality \cite{dohm2018,dohm2019}, leads to exact predictions for all weakly anisotropic systems with periodic boundary  conditions (PBC) in the ($d=2,n=1$) universality class.
We discover unexpected self-similar structures in the critical Casimir amplitude near the instability where weak anisotropy breaks down. They can be traced back to the modular invariance of isotropic CFT.
We also demonstrate the validity of multiparameter universality for confined systems.
More generally, we find self-similar structures
in
the $O(n)$-symmetric $\varphi^4$ theory with PBC
for $1\leq n \leq \infty$ in $d > 2$ dimensions in the presence of planar anisotropies
not only near $T_c$ but also
in the Goldstone-dominated low-temperature region of anisotropic systems with $2\leq n \leq \infty$.

We consider systems with short-range interactions in a rectangular $L_\parallel^{d-1} \times L$  geometry with PBC near an ordinary critical point. The total free energy
${\cal F}_{\rm{tot}}$ (divided by $k_BT$) can be decomposed into singular and nonsingular parts. We are interested in the singular part ${\cal F}_c$ of ${\cal F}_{\rm{tot}}$ at $T_c$. It is well known that the critical free-energy density $f_c={\cal F}_c/(L_\parallel^{d-1} L)$ has the large-$L$ behavior $f_c(L_\parallel,L)=L^{-d}F_c(\rho)$ at fixed aspect ratio $\rho=L/L_\parallel$ \cite{priv,pri}
with a finite amplitude $F_c(\rho)$, which implies that
\begin{eqnarray}
\label{Fcs}
 {\cal F}_c=\rho^{1-d}F_c(\rho)
\end{eqnarray}
is a finite quantity in the large-$L$ limit. The critical Casimir force in the vertical direction is obtained as
\begin{equation}
\label{Fcas}
F_{\mathrm{Cas},c} =  - \partial(L
f_c )/\partial L=L^{-d}X_c(\rho),
\end{equation}
where the derivative is taken at fixed $L_\parallel$. This  yields
the critical Casimir amplitude \cite{dohm2009,dohm2018}
\begin{eqnarray}
\label{Xcas}
X_c(\rho)= (d-1)F_c(\rho)-\rho
 \;\partial F_c(\rho)/\partial\rho = -\rho^d \partial {\cal F}_c/\partial \rho.
\end{eqnarray}
If two-scale-factor universality \cite{pri,priv,privman1990} is valid the amplitudes ${\cal F}_c$, $F_c$, and $X_c$, for given geometry and BC, are universal.
In this Letter we show that these amplitudes
exhibit  a nonuniversal dependence on microscopic couplings with a complex self-similar structure
if the systems are anisotropic.
From CFT we derive
exact
results for $d=2$
for both the scalar $\varphi^4$ model and the Ising model  which belong to the same universality class.

\begin{figure}[t!]
\begin{center}
\includegraphics[width=\columnwidth]{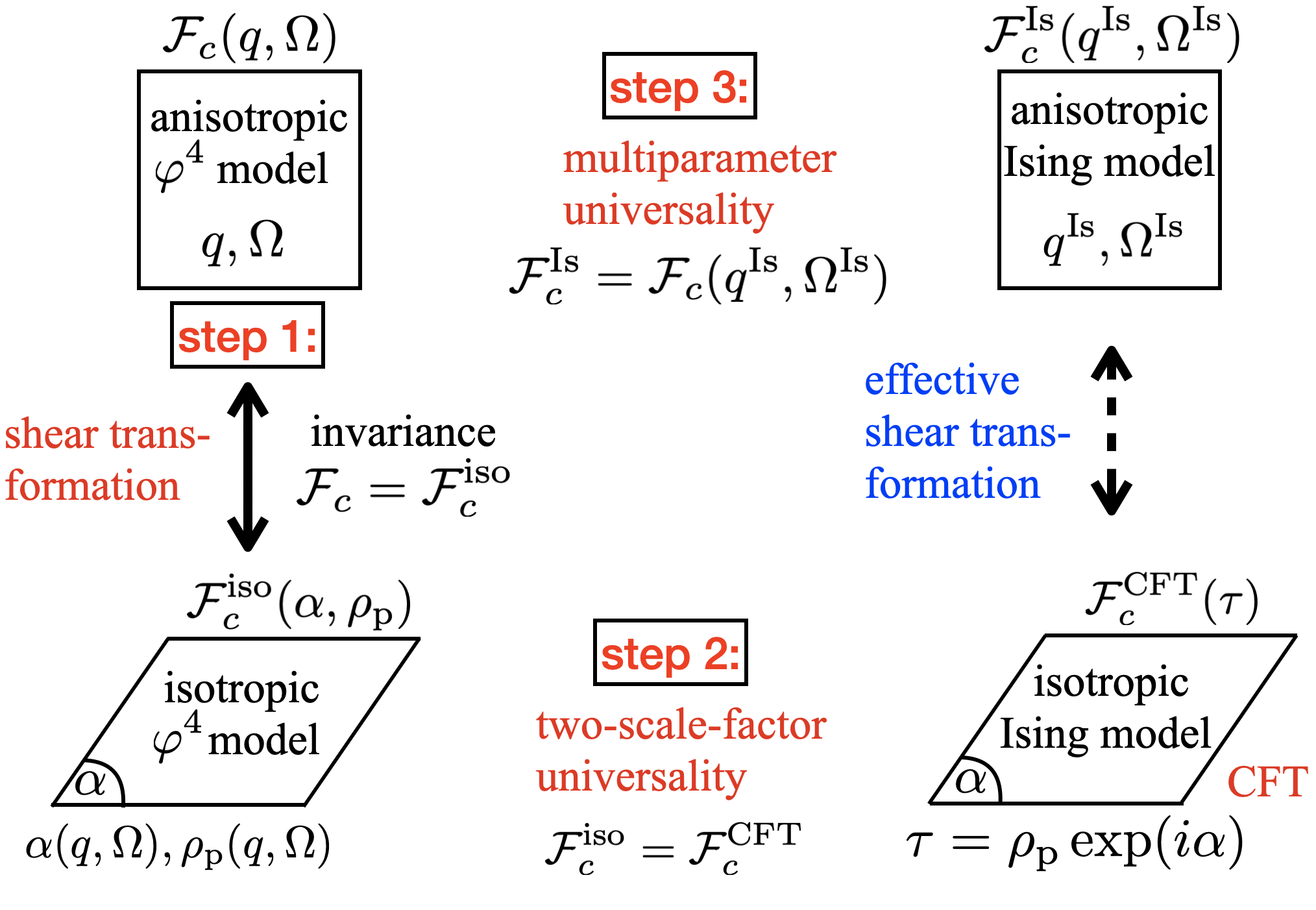}
\end{center}
\caption{Steps of argumentation for the exact relationships between the critical free energies ${\cal F}_c$ of finite anisotropic and isotropic $\varphi^4$  and Ising models in  $d=2$ (see also Fig.~\ref{fig:shear}).}
\label{fig:strategy}
\end{figure}

\begin{figure}[t!]
\begin{center}
\includegraphics[width=\columnwidth]{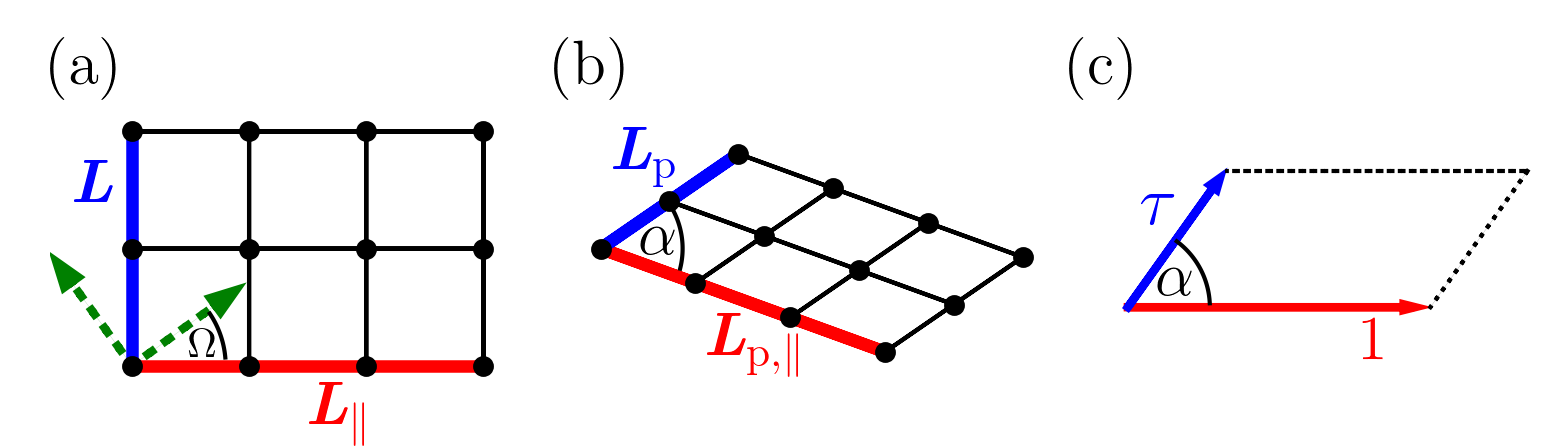}
\end{center}
\caption{
(a) Square lattice with vertical (horizontal) side $\bm{L}$ ($\bm{L}_\parallel$) and aspect ratio $\rho=|\bm{L}|/|\bm{L}_\parallel|$.  Arrows represent the unit vectors ${\bf e}^{(1)}, {\bf e}^{(2)}$ of the principal axes for $\Omega=\pi/5$. (b) Transformed lattice for
$q=1/2$ with sides $\bm{L}_{\rm{p}}$, $\bm{L}_{\rm{p}\parallel}$, and  the angle $\alpha$ and aspect ratio $\rho_{\rm{p}}=|\bm{L}_{\rm{p}}|/|\bm{L}_{\rm{p}\parallel}|$. (c) Parallelogram parametrized by the complex
modular parameter $\tau$, (\ref{tau}).
}
\label{fig:shear}
\end{figure}

We outline our strategy in the schematic Fig.~\ref{fig:strategy} for the case $\rho=1$. The anisotropic $\varphi^4$ model is characterized by two important nonuniversal parameters
(see also Fig.~\ref{fig:shear}):
the angle $\Omega$ describing the orientation of the two principal axes and the ratio $q=\xi_{0\pm}^{(1)}/\xi_{0\pm}^{(2)}$ of the two principal correlation lengths \cite{dohm2019}
$\xi_\pm^{(\beta)}=\xi_{0\pm}^{(\beta)}|t|^{-1}, \beta=1,2$,
$t=(T-T_c)/T_c$. For the anisotropic Ising model the corresponding parameters are denoted by
$\Omega^\mathrm{Is}$ and $q^\mathrm{Is}$.
Step 1 uses a shear transformation of the anisotropic $\varphi^4$  model on a square to an isotropic $\varphi^4$  model on a parallelogram that leaves the critical free energy ${\cal F}_c$ invariant \cite{dohm2006}. Step 2 is based on two-scale-factor universality \cite{priv} implying that the critical free energy ${\cal F}_c^\mathrm{iso}$ of the isotropic $\varphi^4$ model is the same as ${\cal F}_c^{\rm{CFT}}$ of  the isotropic Ising model on the same parallelogram described by CFT. Step 3 employs the hypothesis of multipararmeter universality \cite{dohm2018} predicting that ${\cal F}^\mathrm{Is}_c$ of the anisotropic Ising model with $\rho=1$ is obtained from ${\cal F}_c(q,\Omega)$ of the anisotropic $\varphi^4$ model by the substitution $q \to q^\mathrm{Is}, \Omega \to \Omega^\mathrm{Is}$. Overall, these steps are equivalent to an effective shear transformation (dashed arrow in Fig.~\ref{fig:strategy}) between the isotropic Ising model on a parallelogram and the anisotropic Ising model on a square.

${\it Step}$ 1: We first consider the anisotropic scalar $\varphi^4$ lattice model in a $L_\parallel \times L$ rectangle on a square lattice with lattice spacing $\tilde a$ and PBC. The Hamiltonian and the total free energy divided by $k_B T$
on $N=L_\parallel L/\tilde a^2$ lattice points ${\bf x}_i \equiv (x_{i1}, x_{i2})$ are defined by \cite{dohm2006,dohm2008}
\begin{eqnarray}
\label{2a} H\!& =&\! \tilde a^2  \Big[\sum_{i} \left(\frac{r_0}{2}
\varphi_i^2 + u_0 \varphi_i^4 \right)
 +\sum_{i, j}\!\! \frac{K_{i,j}} {2} (\varphi_i -
\varphi_j)^2\Big] \nonumber
\end{eqnarray}
and by ${\cal F}_{\rm{tot}}=  - \ln \{ \prod_{i = 1}^N \int_{-\infty}^\infty d \varphi_i  \exp \left(- H  \right)\}$.
The large-distance anisotropy is described by the symmetric
anisotropy matrix ${\bf A}=({\bf A}_{\alpha\beta})
=
\left(\begin{array}{ccc}
 a & c \\
 c & b \\
\end{array}\right)$,
\begin{eqnarray}
\label{2i} {\bf A}_{\alpha\beta} &=&\lim_{N \to \infty} N^{-1} \sum_{i,
j} (x_{i \alpha} - x_{j \alpha}) (x_{i \beta} - x_{j \beta})
K_{i,j} .\;\;\;\;
\end{eqnarray}
Weak anisotropy requires positive eigenvalues $\lambda_1>0,\lambda_2>0$ of ${\bf A}$, i.e., $\det {\bf A}>0$ ~\cite{footnote}.
It is known ~\cite{cd2004,dohm2006,dohm2008,kastening-dohm,dohm2018,dohm2019,chen-zhang} that anisotropy effects near $T_c$ are described by the reduced anisotropy matrix  ${\bf \bar A}= {\bf A}/(\det {\bf A})^{1/d}$ which for $d=2$ has the form \cite{dohm2019}
\begin{eqnarray}
\label{Aquer}
{\bf \bar A}(q,\Omega)=
\left(\begin{array}{ccc}
 q \;c_\Omega^2+q^{-1}s_\Omega^2 & \;\;\;(q-q^{-1})\;c_\Omega \;s_\Omega\\
(q-q^{-1})\; c_\Omega\; s_\Omega& q \;s_\Omega^2 +q^{-1}\;c_\Omega^2
\end{array}\right)
\end{eqnarray}
with $q=(\lambda_1/\lambda_2)^{1/2}=\xi_{0\pm}^{(1)}/\xi_{0\pm}^{(2)}$ and the abbreviations $c_\Omega\equiv\cos\Omega,s_\Omega\equiv\sin\Omega$ where $\Omega$ determines the  principal axes described by the eigenvectors
${\bf e}^{(1)}=(c_\Omega,s_\Omega)^\intercal$, ${\bf e}^{(2)}=(-s_\Omega,c_\Omega)^\intercal$ of ${\bf A}$.
The exact dependence of $\Omega$ and $q$ on the couplings $K_{i,j}$ through $a,b,c$ has been derived in \cite{dohm2019}.
A shear transformation
can be performed such that the transformed $\varphi^4$ model on a parallelogram (Fig.~\ref{fig:shear}) has changed second moments ${\bf A}'_{\alpha\beta}= \delta_{\alpha \beta}$
representing an isotropic system
\cite{cd2004,dohm2008,dohm2019,kim1987}.
The transformations  $\bm{L}_{\rm{p}}={\bm \lambda}^{-1/2}{\bf U}{\bm L}$ and $\bm{L}_{\rm{p}\parallel}={\bm \lambda}^{-1/2}{\bf U}{\bm L}_\parallel$ of the vertical and horizontal sides  $\bm L=L{\bf e}_v$ and  $\bm L_\parallel=L_\parallel   {\bf e}_h$
yield the corresponding transformed sides $\bm{L}_{\rm{p}}$ and $\bm{L}_{\rm{p}\parallel}$ of the parallelogram where the rotation and rescaling matrices
\begin{eqnarray}
&&{\bf{ U}}=\left(\begin{array}{ccc}
 \cos\; \Omega& \sin\; \Omega \\
  -\sin\; \Omega &  \cos\; \Omega \\
\end{array}\right),\;\;\;\;
{{\bm \lambda}}
=\left(\begin{array}{ccc}
 \lambda_1& 0 \\
 0 & \;\lambda_2 \\
\end{array}\right)\;\;\;\;\;\;\;
\end{eqnarray}
are employed.
The parallelogram is characterized by the angle $0<\alpha< \pi$ between $\bm{L}_{\rm{p}}$ and $\bm{L}_{\rm{p}\parallel}$ and the transformed aspect ratio $\rho_{\rm{p}}= |\bm{L}_{\rm{p}}|/|\bm{L}_{\rm{p}\parallel}|$.
We find
\begin{eqnarray}
\label{alpha}
&& \cot \alpha(q,\Omega)
= -{{\bf \bar A}(q,\Omega)}_{12}= (q^{-1}-q)\cos \Omega \;\sin \Omega,\;\;\;\;\;\;\;\;
\\
\label{lengthratio}
&& [\rho_{\rm{p}}(\rho,q,\Omega)]^2= \rho^2\; \frac{{\bf \bar A}(q,\Omega)_{11}}{{\bf \bar A}(q,\Omega)_{22}}=\rho^2\; \frac{\tan^2\Omega+q^2}{1+q^2\tan^2\Omega},\;\;\;\;\;\;\;
\end{eqnarray}
for arbitrary $\rho,q,\Omega$ which is valid
for arbitrary BC.
The singular part ${\cal F}^{\rm{iso}}_{c\rm{}}(\alpha,\rho_{\rm{p}})$ of the total free energy at $T_c$ of the isotropic parallelogram is a function of $\alpha$ and $\rho_{\rm{p}}$.
The shear transformation leaves both the Hamiltonian
and the singular part ${\cal F}_c$ of the total free energy
${\cal F}_{\rm{tot}}$ at $T_c$ invariant \cite{dohm2006,dohm2008}, thus  ${\cal F}_c$ is determined by
\begin{eqnarray}
\label{freeEnergyanisox}
{\cal F}_c(\rho,q,\Omega)
={\cal F}^{\rm{iso}}_{c}\big(\alpha(q,\Omega),\rho_{\rm{p}}(\rho,q,\Omega)\big)
=\rho^{-1}F_c(\rho,q,\Omega).
\end{eqnarray}
In the strip limit the shear transformation yields
~\cite{suppl}
\begin{eqnarray}
\label{Casnull}
\lim_{\rho\to 0} X_c(\rho,q,\Omega)=
(q \cos^2\Omega+q^{-1}\sin^2\Omega)^{-1}X^{\rm{iso}}_{c,\rm{strip}},
\end{eqnarray}
where $X^{\rm{iso}}_{c,\rm{strip}}$ is the amplitude on an isotropic strip.
Eqs. (\ref{freeEnergyanisox}) and (\ref{Casnull}) demonstrate that ${\cal F}_c$, $F_c$,and  $X_c$
depend on microscopic details via $q(a,b,c)$ and $\Omega(a,b,c)$, thus violating two-scale-factor universality.
So far
it is unknown how to calculate the
dependence of ${\cal F}^{\rm{iso}}_{c}$  on $\alpha$ and $\rho_{\rm{p}}$.

${\it Step}$ 2: At this point we invoke two-scale-factor universality for {\it isotropic} systems \cite{pri,dohm2018} which means that isotropic  $\varphi^4$ and Ising models
have the same singular parts ${\cal F}^{\rm{iso}}_{c\rm{}}$ and ${\cal F}^{\rm{Is},iso}_{c\rm{}}$.
For the Ising model exact information is available  from CFT \cite{franc1987,franc1997}.
Via
an isotropic continuum description in terms of a free fermion field
an exact
contribution $Z^{\rm{CFT}}(\tau)$ to the partition function of the $d=2$ isotropic Ising model on a torus at $T_c$ has been derived.
We choose the same
parameters $\alpha$ and $\rho_{\rm{p}}$ as for the
isotropic $\varphi^4$ model.
The Ising parallelogram is described by a complex torus modular parameter \cite{franc1997}
\begin{eqnarray}
\label{tau}
\tau(\alpha,\rho_{\rm{p}}) = \mbox{Re}\; \tau + i\; \mbox{Im} \;\tau=\rho_{\rm{p}}\exp(i\; \alpha)
\end{eqnarray}
where $ \alpha$ is the angle shown in Fig.~\ref{fig:shear}(c) and $\rho_{\rm{p}}=|\tau|$ is the aspect ratio of the Ising parallelogram. The partition function is expressed in terms of Jacobi theta functions $\theta_i(0|\tau)\equiv \theta_i(\tau)$ (in the notation of \cite{franc1997}, see \cite{suppl}) as~\cite{franc1987}
\begin{equation}
\label{ZIsing}
Z^{\rm{CFT}}(\tau)=\big({|\theta_2(\tau)|+|\theta_3(\tau)|+|\theta_4(\tau)|}\big)/\big({2|\eta(\tau)|}\big),
\end{equation}
with
$\eta(\tau)=(\frac{1}{2}\theta_2(\tau)\theta_3(\tau)\theta_4(\tau))^{1/3}$, from which we obtain
${\cal F}^{\rm{CFT}}_{c}(\tau)=\!-\ln Z^{\rm{CFT}}(\tau).$
%
The singular part of the total free energy of the isotropic Ising model at $T_c$ is
\begin{eqnarray}
\label{freeCFT}
{\cal F}^{\rm{Is},iso}_{c\rm{}}\!\big(\tau(\alpha,\rho_{\rm{p}})\big)
\!={\cal F}^{\rm{CFT}}_c\!\big(\tau(\alpha,\rho_{\rm{p}})\big)
\!= {\cal F}^{\rm{iso}}_{c\rm{}}\!\big(\alpha,\rho_{\rm{p}}\big),
\end{eqnarray}
where, owing to two-scale-factor universality, the last equation applies to the transformed $\varphi^4$ model on the  parallelogram. We define $\tau(\rho,q,\Omega)= \tau\big(\alpha(q,\Omega),\rho_{\rm{p}}(\rho,q,\Omega)\big)$
with $\alpha(q,\Omega)$ and $\rho_{\rm{p}}(\rho,q,\Omega)$ given by (\ref{alpha}) and (\ref{lengthratio}). Then we obtain from (\ref{freeCFT}), (\ref{freeEnergyanisox}), (\ref{Fcs}), and (\ref{Xcas}) our exact result for the Casimir amplitude $X_c$ of the anisotropic $\varphi^4$ model
as
\begin{eqnarray}
\label{FcasCFT}
X_c(\rho,q,\Omega)&=& -\rho^2 \;\partial {\cal F}_c(\rho,q,\Omega)/\partial\rho
\end{eqnarray}
with
${\cal F}_c(\rho,q,\Omega)={\cal F}_c^{\rm{CFT}}\big(\tau(\rho,q,\Omega)\big)$
where the  nonuniversal expressions for $q(a,b,c)$ and $\Omega(a,b,c)$
\cite{dohm2019}
have to be inserted.
In the strip limit we obtain ~\cite{suppl}
\begin{eqnarray}
\label{FcasCFTstrip}
X_c(0,q,\Omega)= - \pi/[12(q \cos^2\Omega+q^{-1}\sin^2\Omega)],
\end{eqnarray}
%
in accord with the CFT result $- \pi/12$
 ~\cite{bloete,affleck} for $q=1$.

%
\begin{figure}[t!]
\begin{center}
\includegraphics[height=3.4cm,keepaspectratio,]{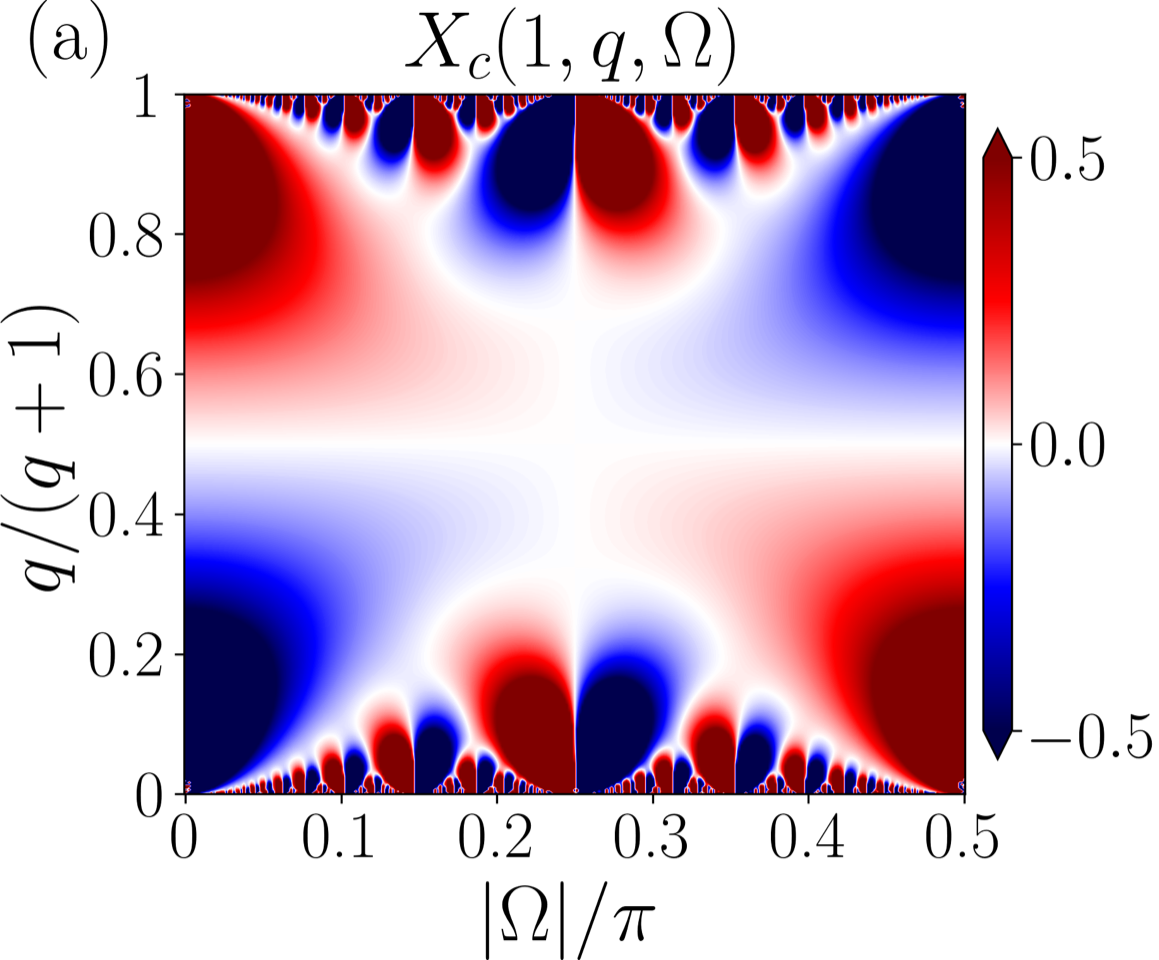}
\includegraphics[height=3.4cm,keepaspectratio,]{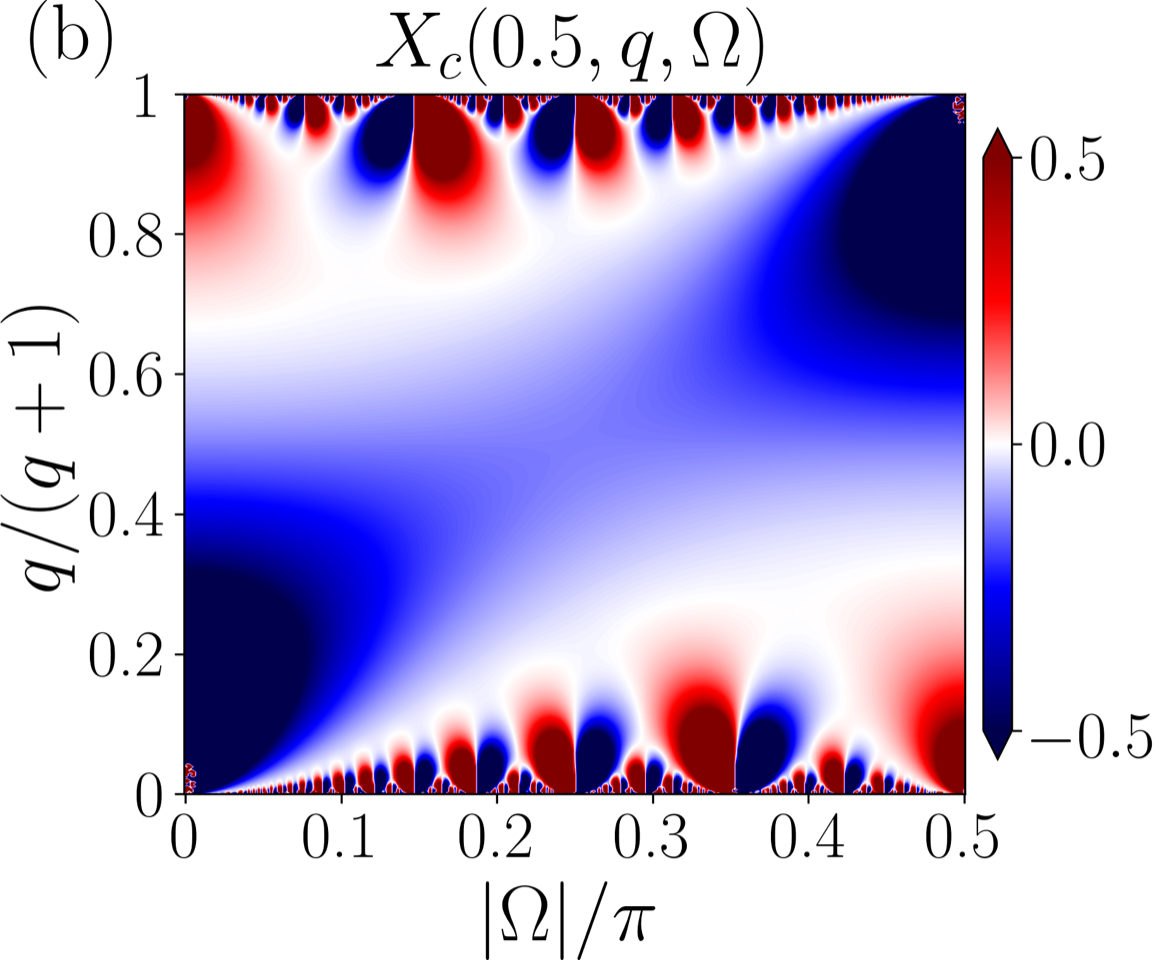}
\end{center}
\caption {Universal contour plot of the critical Casimir amplitude $X_c$ of the anisotropic $d=2$ $\varphi^4$ model from (\ref{FcasCFT}) for $\rho=1$ in (a) and for $\rho=0.5$ in (b).  For the anisotropic $d=2$  Ising model, (\ref{FcasCFTIsing}) yields the same plots if $q\to q^\mathrm{Is}, \Omega\to \Omega^\mathrm{Is}$.
}
\label{fig:universal}
\end{figure}
\begin{figure}[t!]
\begin{center}
\includegraphics[width=0.495\columnwidth]{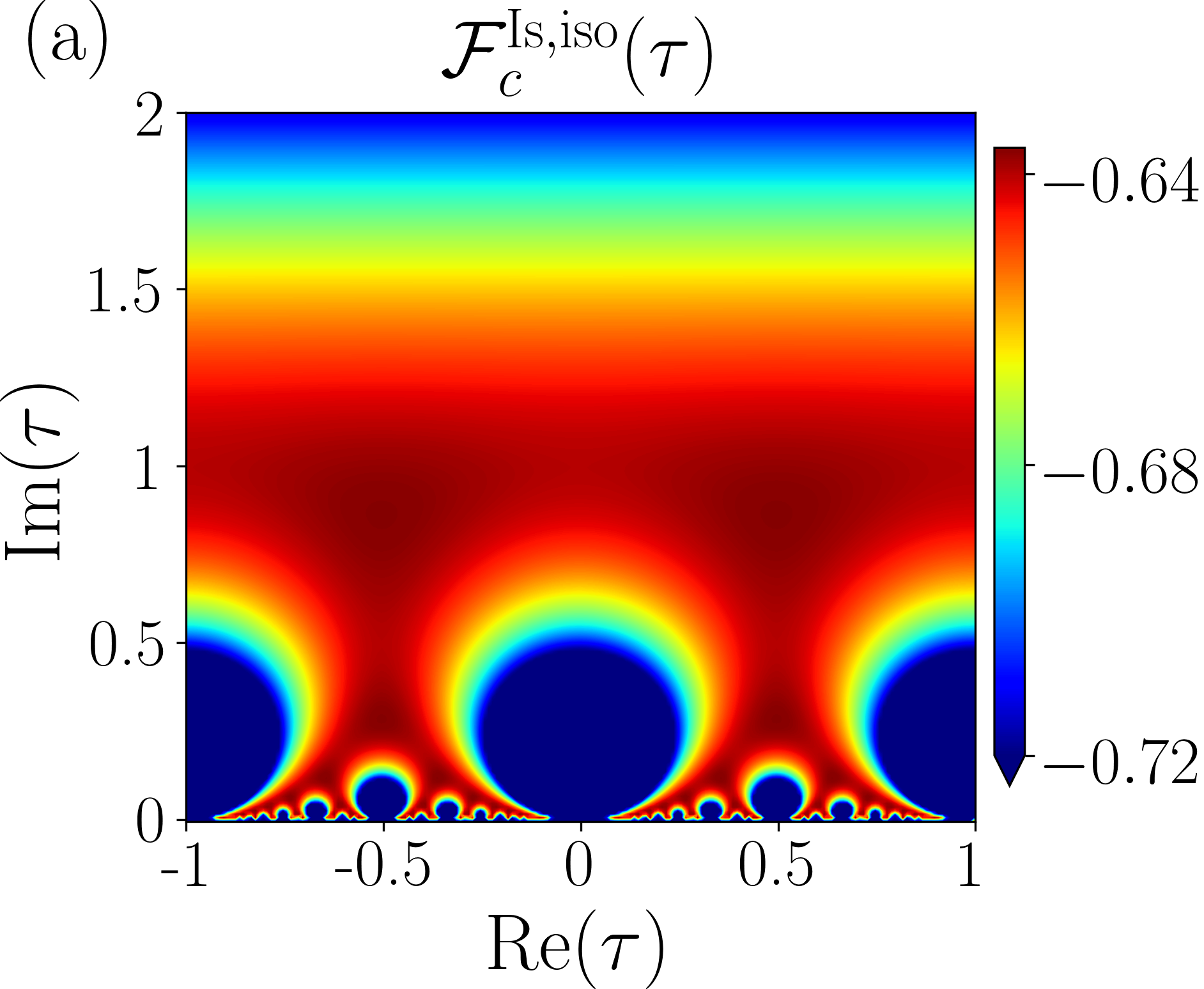}
\includegraphics[width=0.49\columnwidth]{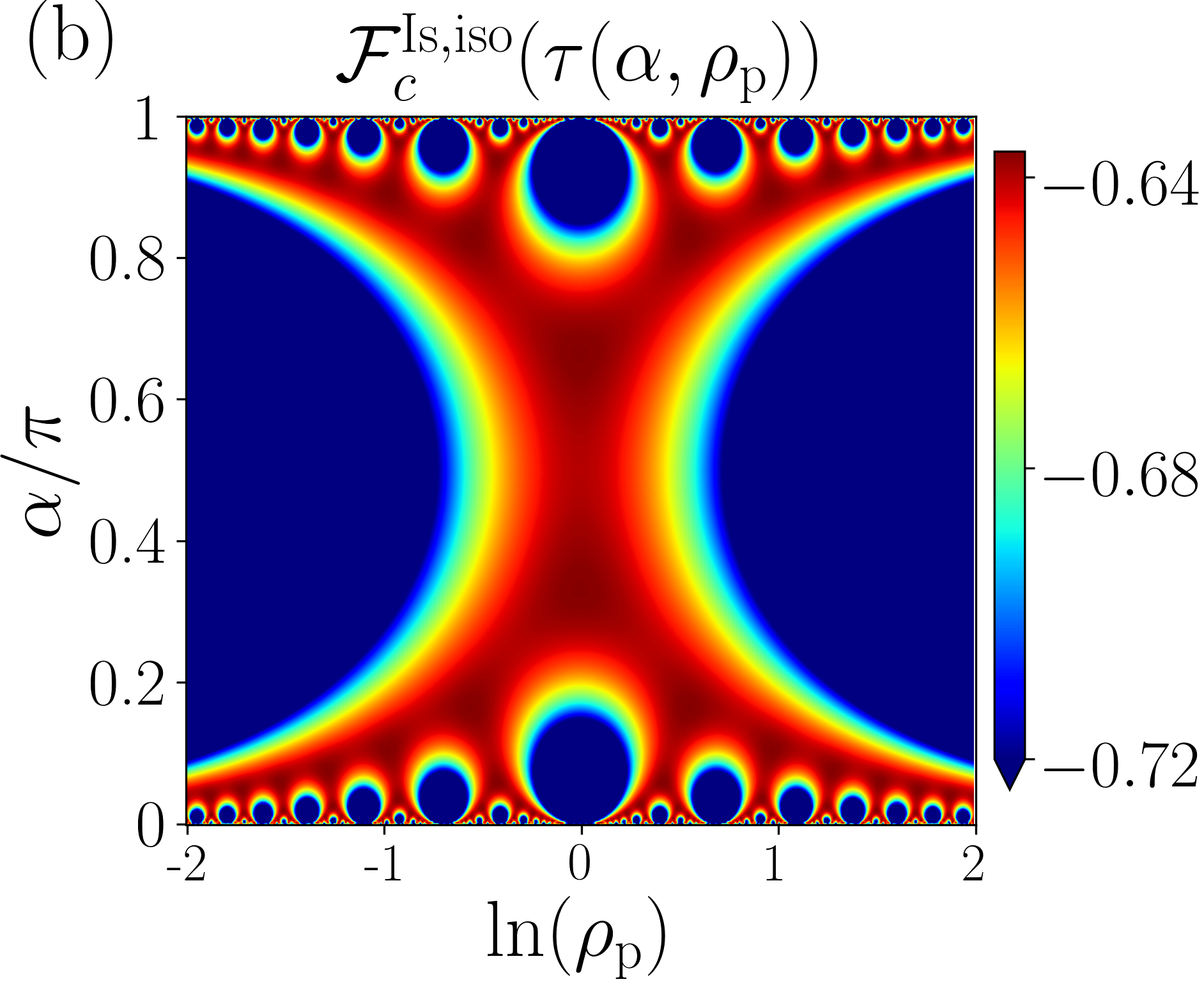}
\end{center}
\caption{Critical amplitude ${\cal F}^{\rm{Is},iso}_{c\rm{}}$, (\ref{freeCFT}), of the isotropic Ising model in parallelogram geometry (a) in the complex $\tau$ plane and (b) in the $\rho_p - \alpha$ plane. The same plot (b)  holds for ${\cal F}^{\rm{iso}}_{c\rm{}}$, (\ref{freeCFT}), of the transformed isotropic $\varphi^4$ model.
}
\label{fig:CFT}
\end{figure}

In Fig.~\ref{fig:universal} we
present contour plots of $X_c$ in the complete $\Omega-q$ plane
for $\rho=1$ and $\rho=0.5$. Contrary to the simple ($\Omega,q$) dependence (\ref{FcasCFTstrip}) in strip geometry and to the claim that the effects of weak anisotropy are fairly harmless \cite{diehl2010}, we find unexpectedly complex structures exhibiting the feature of self-similarity in the regions $ q\ll 1$ and $q\gg 1$ near the border lines $q=0$ and $q=\infty$, where  $\det {\bf A} = 0$ or
$\lambda_\alpha = 0$, i.e., where weak anisotropy breaks down  \cite{footnote}.
This self-similarity can be traced back to the property of modular invariance \cite{franc1997} $Z^{\rm{CFT}}(\tau)= Z^{\rm{CFT}}(\tau+1)$, $Z^{\rm{CFT}}(\tau)= Z^{\rm{CFT}}(- 1/\tau)$  for the partition function of the isotropic Ising model at $T_c$ in a parallelogram geometry with PBC, i.e., on a torus,  which implies ${\cal F}^{\rm{Is},iso}_{c\rm{}}(\tau)={\cal F}^{\rm{Is},iso}_{c\rm{}}(- 1/\tau)={\cal F}^{\rm{Is},iso}_{c\rm{}}(\tau+1).$
This is illustrated by the periodic structure of ${\cal F}^{\rm{Is},iso}_{c\rm{}}$ in the complex $\tau$ plane [Fig.~\ref{fig:CFT}(a)]
which generates a self-similar structure in the $(\rho_{\rm{p}},\alpha)$ plane [Fig.~\ref{fig:CFT}(b)].
The modular transformation $\tau \to -1/\tau$ corresponds to $\rho_{\rm{p}} \to 1/\rho_{\rm{p}}, \alpha \to \pi-\alpha$ which yields equivalent parallelograms.
The Dehn twist   \cite{franc1997} $\tau \to \tau +1$ yields parallelograms with different
$\rho_{\rm{p}}$ and $\alpha$,
but the invariance of ${\cal F}^{\rm{Is},iso}_{c\rm{}}$
can be understood geometrically since a given torus can be cut in different ways such that different parallelograms with PBC are generated which all have the same critical free energy
on the same torus. By two-scale-factor universality, the same result applies to ${\cal F}^{\rm{\rm{iso}}}_{c\rm{}}$ of the isotropic $\varphi^4$ model on the same torus. The dependence on ($\rho_{\rm{p}},\alpha$) for the isotropic system in Figs.~\ref{fig:shear}(b) and \ref{fig:CFT}(b) is transferred by the shear transformation
(\ref{alpha}), (\ref{lengthratio}) and by (\ref{FcasCFT})
to a corresponding dependence of $X_c(\rho, q,\Omega)$ on $(q,\Omega)$ as is shown in Fig.~\ref{fig:universal}.
We note that so far no assumption has been made other than the validity of two-scale-factor universality for isotropic systems.

\begin{figure}[t!]
\begin{center}
\end{center}
\includegraphics[width=0.49\columnwidth]{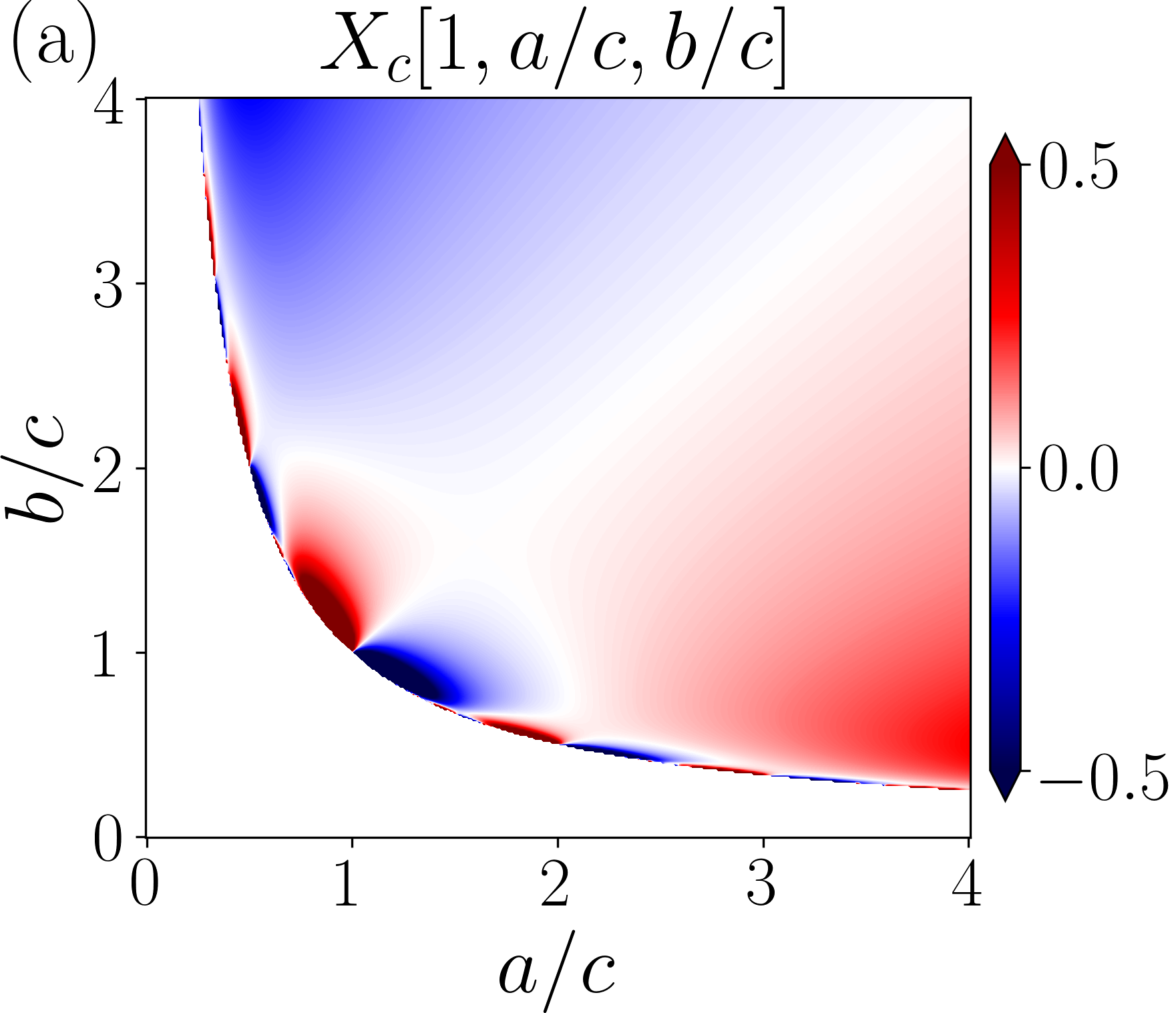}
\includegraphics[width=0.49\columnwidth]{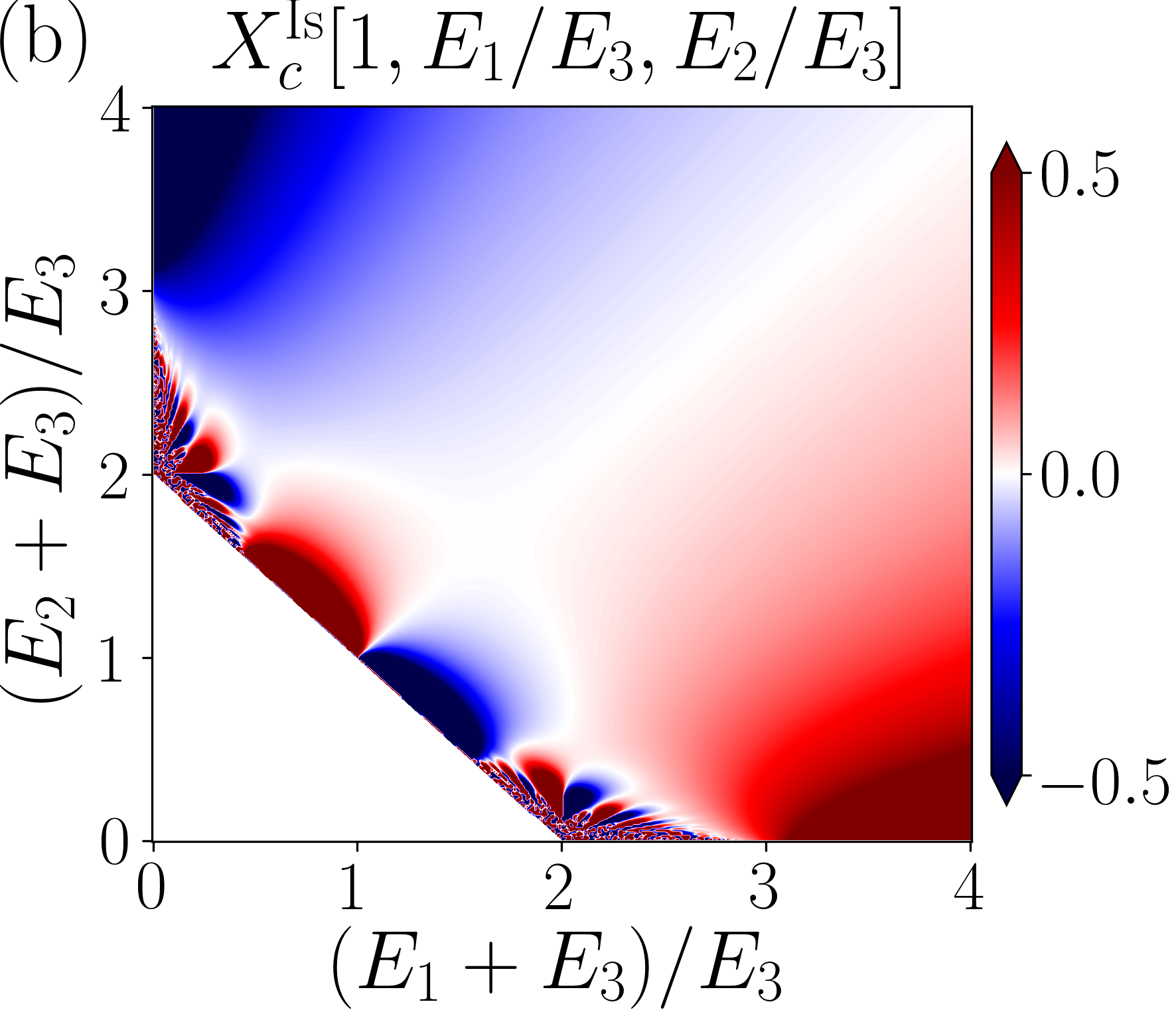}
\caption{
Nonuniversal critical Casimir amplitudes $X_c$ and $X^{\rm{Is}}_c$ for $\rho=1$ of the  $d=2$ $\varphi^4$ and Ising models for (a) $\det {\bf A} =ab-c^2 > 0$, (b)  $E_1+E_2>0, E_1+E_3>0, E_2+E_3>0$ \cite{Berker}.
}
\label{fig:phi4Ising}
\end{figure}

${\it Step}$ 3: We proceed to the
anisotropic triangular Ising  model on an $L_\parallel \times L$ rectangle with  the Hamiltonian ~\cite{dohm2019,Indekeu}
%
$H^{{\text{\rm{Is}}}}\! =\!-\!\!\sum_{j, k}\big [E_1\sigma_{j,k} \sigma_{j,k+1}+E_2\sigma_{j,k} \sigma_{j+1,k}  +E_3\sigma_{j,k} \sigma_{j+1,k+1}\big]$
%
with spin variables $\sigma_{j,k}=\pm1$   on a square lattice
with PBC. Both the  angle
$\Omega^{\rm{Is}}(E_1,E_2,E_3)$ of the principal axes and the  ratio of the principal correlation lengths $q^{\rm{Is}}(E_1,E_2,E_3)=\xi_{0\pm}^{(1)\rm{Is}}/\xi_{0\pm}^{(2)\rm{Is}}$  are known
functions of
$E_1,E_2,E_3$ \cite{dohm2019}.
Multiparameter universality
was proven for bulk systems in \cite{dohm2019}, thus the exact critical bulk correlation function is governed by the Ising anisotropy matrix
${\bf \bar A}^{\rm{Is}}={\bf \bar A}(q^{\rm{Is}},\Omega^{\rm{Is}})$
with the same matrix ${\bf \bar A}$ as in (\ref{Aquer}) for the $\varphi^4$ model, but  with $q$ and $\Omega$ replaced by $q^{\rm{Is}}$ and $\Omega^{\rm{Is}}$. Since bulk and finite-size properties are governed by the same anisotropy matrix~\cite{dohm2018} we
predict that
the exact critical Casimir amplitude
of the anisotropic Ising model
is given by
\begin{eqnarray}
\label{FcasCFTIsing}
&&X^{\rm{Is}}_c(\rho,q^{\rm{Is}},\Omega^{\rm{Is}}) =  -\rho^2 \;\partial {\cal F}_c(\rho,q^{\rm{Is}},\Omega^{\rm{Is}})/\partial\rho,
\end{eqnarray}
where
${\cal F}_c(\rho,q^{\rm{Is}},\Omega^{\rm{Is}})$
is the same function as in (\ref{FcasCFT}) 
but now
the
results for $q^{\rm{Is}}(E_1,E_2,E_3)$ and $\Omega^{\rm{Is}}(E_1,E_2,E_3)$ of the Ising model \cite{dohm2019} have to be inserted. Our predictions go far beyond all previous special results \cite{bloete,affleck,rud2010,FF,salas,izmail,kastening2012,HH2019} for confined isotropic and anisotropic Ising models. Here we have succeeded in treating the general anisotropic case of an arbitrary direction of the principal axes described by a nonzero angle $\Omega^{\rm{Is}}$ in a finite geometry with an arbitrary aspect ratio $\rho$. This is of physical relevance for general anisotropic systems with more complicated interactions whose principal axes generically have skew directions relative to the symmetry axes of the underlying lattice. This advance is made possible by our new approach of combining exact relations of anisotropic $\varphi^4$ lattice theory with exact results of CFT. Specifically our
predictions agree with Ising-model results for isotropic strips \cite{bloete,affleck,rud2010}, rectangles \cite{FF}, and parallelograms \cite{salas} as well as for anisotropic strips \cite{kastening2012} and rectangles \cite{izmail,HH2019} which
constitutes a direct confirmation of
multiparameter universality for confined systems.
Thus we predict that the results in  Fig.~\ref{fig:universal} for the 
$\varphi^4$ model
are valid also for the
Ising  model after substituting $q\to q^\mathrm{Is}, \Omega\to \Omega^\mathrm{Is}$, i.e., the $(q,\Omega)$  representation has a universal character that is applicable to all weakly anisotropic systems in the ($d=2,n=1$) universality class. It becomes nonuniversal if the dependence of $(q,\Omega)$ and $(q^{\rm{Is}},\Omega^{\rm{Is}})$ on $a,b,c$ and $E_1, E_2$, $E_3$
is inserted. We denote these
Casimir amplitudes by   $X_c[\rho,a/c,b/c]$ and  $X^{\rm{Is}}_c[\rho,E_1/E_3,E_2/E_3]$. They are shown in Fig.~\ref{fig:phi4Ising} for $\rho=1$.
The nonuniversal differences between $X_c$ and $X^{\rm{Is}}_c$ confirm the prediction \cite{cd2004,dohm2006,dohm2008,dohm2018} that
the Casimir amplitude $X_c$ for weakly anisotropic systems is not a universal quantity.
Even if it is known for one anisotropic system it cannot be predicted for other anisotropic systems of the same universality class  since  $\Omega$ generically depends in an unknown nonuniversal way on the anisotropic interactions \cite{dohm2019}.
%
%

\begin{figure}[t!]
\begin{center}
\end{center}
\includegraphics[height=3.5cm]{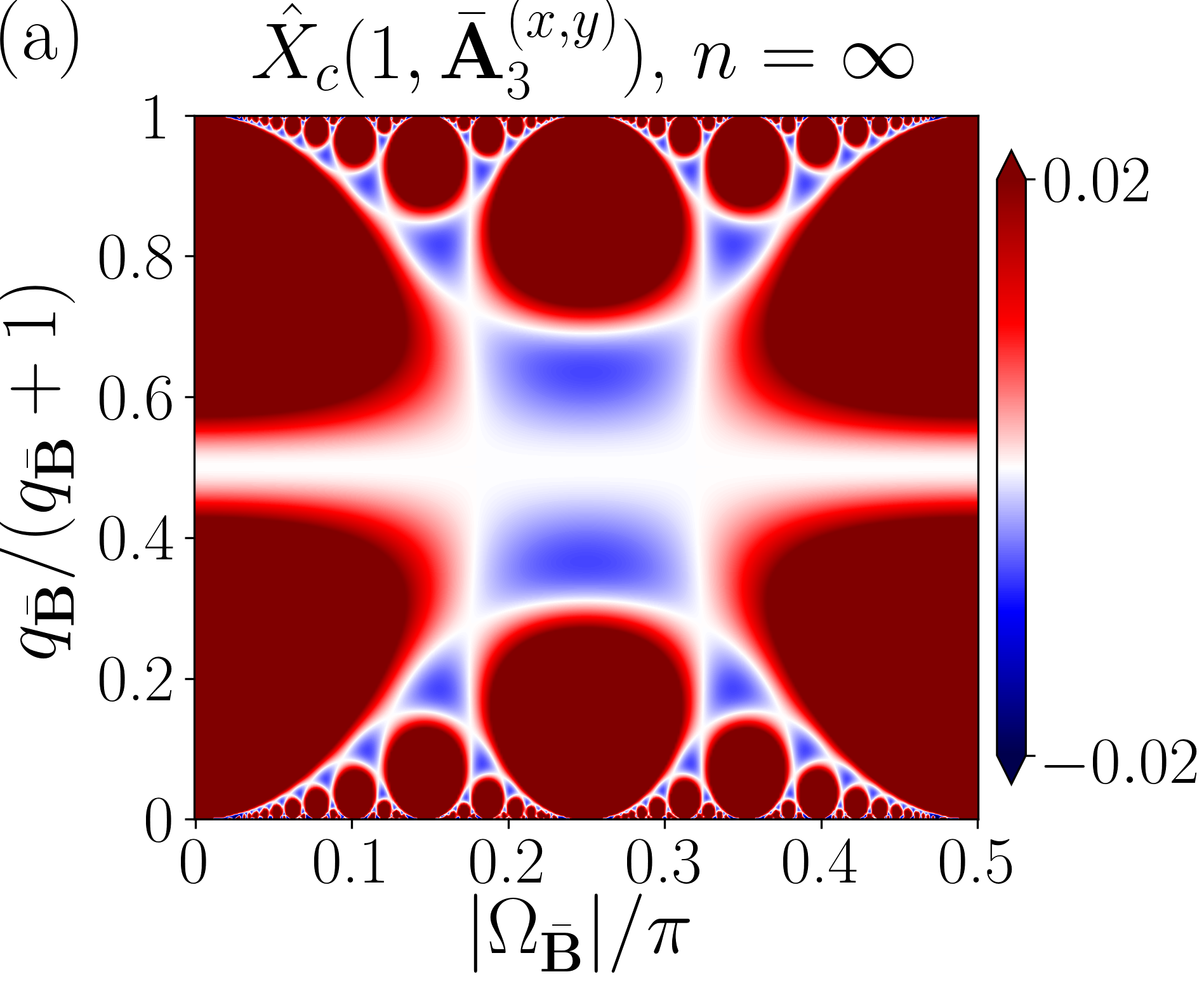}
\includegraphics[height=3.5cm]{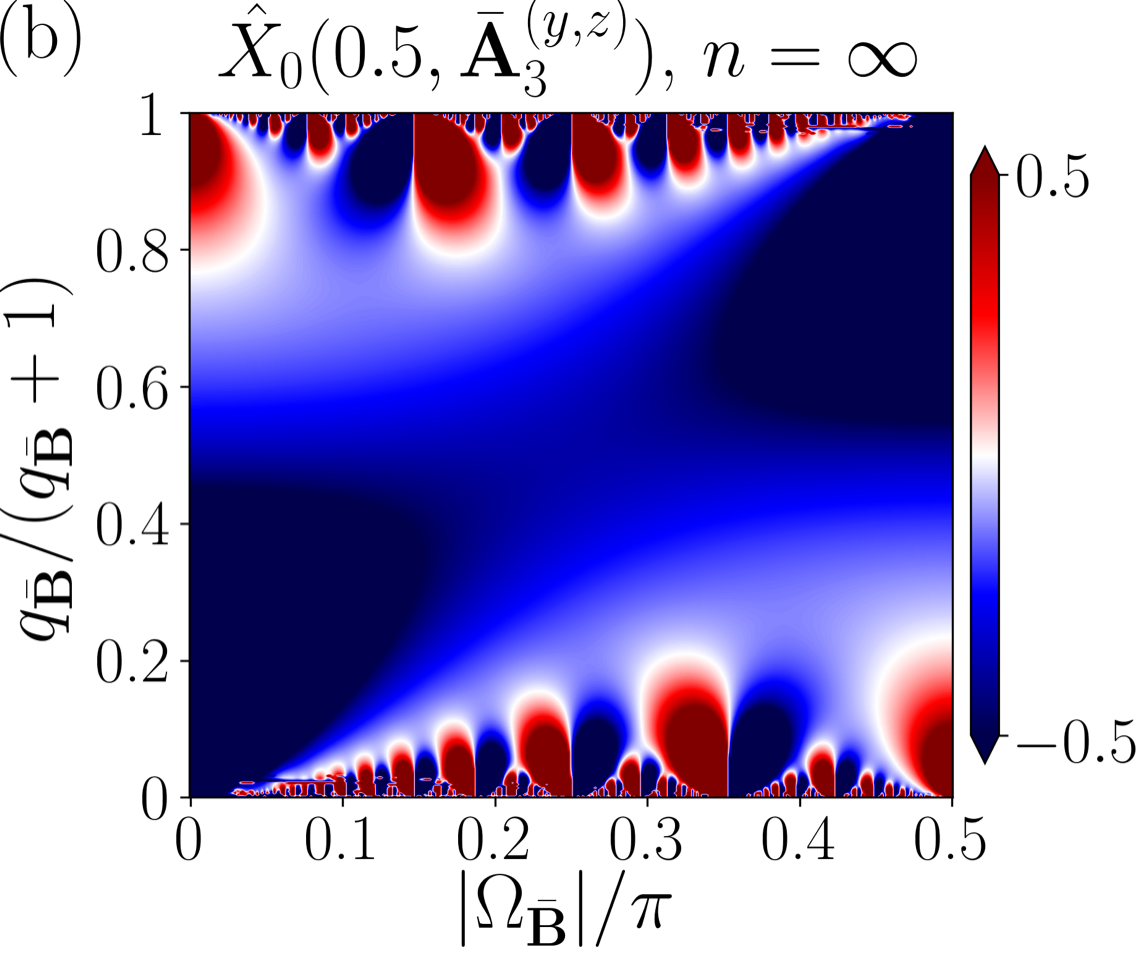}
\caption{Universal contour plots of critical and low-temperature Casimir amplitudes
of the $d=3$ $\varphi^4$ theory for $n=\infty$ \cite{dohm2018,suppl} with planar anisotropies (\ref{redaniso}):
$\hat X_c$ for $\rho=1$  with ${\bf \bar A^{(x,y)}_3}$ in (a) and  $\hat X_0$ for  $\rho=0.5$ with ${\bf \bar A^{(y,z)}_3}$ in (b).
}
\label{fig:3D}
\end{figure}

In the following we demonstrate that self-similar structures exist
more generally for $O(n)$-symmetric systems with PBC for $d=3,n\geq 1$ with a $3 \times 3$ anisotropy matrix ${\bf A_3}$. Consider a
$ L_\parallel^2 \times L$ geometry with $\rho=L/ L_\parallel$.
In \cite{dohm2018} the  scaling function $X(\hat x, \rho,{\bf \bar A_3})$ of the Casimir force of the $\varphi^4$ model has been derived for $n \geq 1$,
where $\hat x\propto (T-T_c)L^{1/\nu}$ is the
scaling variable.  This includes the low-temperature amplitude $X_0(\rho,{\bf \bar A_3})\equiv X(-\infty, \rho,{\bf \bar A_3})$  due to the Goldstone modes for $n\geq 2$. In particular, the exact result $\hat X = \lim_{n \to \infty}\;X/n$
has been derived in the large-$n$ limit. At fixed $\hat x$ the anisotropy effect is completely contained in the function
\begin{eqnarray}
\label{substKxx}
&&K_3 (y, {\bf \widehat C})=\sum_{\bf m} \;\exp (-  y\;
{\bf m} \cdot {\bf \widehat C m})
\end{eqnarray}
where $y$ is independent of ${\bf \bar A_3}$. The sum $\sum_{\bf m}$ runs over ${\bf m} = (m_1, m_2, m_3) \; , m_\alpha = 0, \pm 1,..., \pm \infty$  and the $3\times 3$ matrix ${\bf \widehat C}$ has the elements $\widehat C_{\alpha\beta}=\rho_\alpha \;\rho_\beta \;({\bf \bar A_3})_{\alpha\beta}$
with $\rho_1=\rho_2=\rho,\rho_3=1$.
We consider two types of planar anisotropies as described by the anisotropy matrices
\begin{eqnarray}
 \label{redaniso}
&& {\bf \bar A^{(x,y)}_3}
 = \left(\begin{array}{ccc}
    {\bf \bar B_h} &  0  \\
  0 & { 1}
\end{array}\right),\;\;\;{\bf \bar A^{(y,z)}_3}
 = \left(\begin{array}{ccc}
    { 1} &  0  \\
  0 & {\bf \bar B_v}
\end{array}\right).
\end{eqnarray}
In (\ref{redaniso}) the $2 \times 2$ submatrices ${\bf \bar B_h}$ and ${\bf \bar B_v}$ describe anisotropies in the ``horizontal'' $x-y$ and ``vertical'' $y-z$ planes, respectively. The difference between these cases is that the Casimir force defined in (\ref{Fcas}) is perpendicular to the $ x-y$ anisotropy in case of   ${\bf \bar A^{(x,y)}_3}$ whereas it is parallel to the $y-z$ anisotropy in case of  ${\bf \bar A^{(y,z)}_3}$.
Both  ${\bf \bar B_h}$ and  ${\bf \bar B_v}$ have the same form as
in (\ref{Aquer}), with $(q,\Omega )$ replaced by $(q_{\bf \bar B},\Omega_{\bf \bar B})$.
This suggests that self-similar structures  exist for $d=3$ like those found for $d=2$.
This is indeed verified by evaluating  the exact
results for $\hat X_c$ and $\hat X_0$ of \cite{dohm2018} for $d=3,n=\infty$ with
the planar anisotropies (\ref{redaniso}) as shown in Fig.~\ref{fig:3D}.
 We find similar structures from \cite{dohm2018} for any finite $n$ and $\hat x$.
The self-similar structures of Fig.~\ref{fig:3D} disappear in the film limit $\rho \to 0$ \cite{dohm2018} at finite $L$ but are maintained in the cylinder limit $\rho \to \infty$ \cite{dohm2011} at finite $L_\parallel$.

We find that, to some extent, this self-similarity can be traced back to the
modular invariance of $K_3 (y, {\bf \widehat C})$,
(\ref{substKxx}).
Since a symmetric matrix ${\bf \bar B}$  with $\det {\bf \bar B}=1$ contains only two independent matrix elements, it can be expressed
as
\begin{eqnarray}
\label{barB}
{\bf \bar B}
=
\frac{1}{\mbox{Im}(\tau_{\bf \bar B})}\left(\begin{array}{ccc}
 |\tau_{\bf \bar B}|^2 & -\mbox{Re}( \tau_{\bf \bar B}) \\
 -\mbox{Re}( \tau_{\bf \bar B})& 1 \\
\end{array}\right),
\end{eqnarray}
where $\tau_{\bf \bar B}= \mbox{Re}( \tau_{\bf \bar B}) + i\; \mbox{Im}(\tau_{\bf \bar B}) $ is a complex number with $\mbox{Im}(\tau_{\bf \bar B})>0$.
Based on the one-to-one relation (\ref{barB}) between  ${\bf \bar B}$ and $\tau_{\bf \bar B}$,
we can relate a modular transformation $\tau_{\bf \bar B} \to \tau_{\bf \bar B'}$  to a corresponding matrix ${\bf \bar B}'$, with, e.g.,
$\tau_{\bf \bar B'} =  \tau_{\bf \bar B} +1$ for the Dehn twist.
The function $K_3$ remains invariant under such
transformations for ${\bf \bar A_3}={\bf \bar A^{(y,z)}_3}$, $\rho=1$ and for ${\bf \bar A_3}={\bf \bar A^{(x,y)}_3}$ and arbitrary $\rho$~\cite{suppl}.
This is parallel to the modular invariance of $Z^{\rm{CFT}}(\tau)$.
More generally, we expect
self-similar structures
also for $d=3$
systems with  non-planar
anisotropies
and PBC.

{\it Conclusion and outlook -}
We have studied the
dependence of finite-size effects on the principal correlation lengths and
principal axes for the case of PBC.
For $d=2$,
we have achieved a breakthrough
by identifying unexpected self-similar structures
via the combination of
isotropic CFT with anisotropic $\varphi^4$ theory.
For $d=3$, our analysis paves the way towards an exploration of
finite-size effects near the borderlines where weak anisotropy breaks down not only near $T_c$ but also
in the Goldstone-dominated region.
On the basis of $d=3$ finite-size theories
\cite{Esser,CDS,dohm2018,dohm2008} and owing to multiparameter universality we predict that in all $O(n)$-symmetric systems with weak anisotropies and PBC the self-similar structures described in this Letter appear also in
various physical quantities such as the specific heat
and susceptibility.
Self-similar structures
do not appear for simple anisotropies with $\Omega = 0,\pi/4$ studied previously
\cite{izmail,Selke2009,HH2019,kim1987,Yurishchev} although two-scale-factor universality is violated in these cases. Our results
provide strong motivation for investigating the case of other boundary conditions and to study finite-size effects in
anisotropic systems such as superconductors, magnetic materials, solids with structural phase transitions and near magnetic-field-induced phase transitions \cite{lin} where the interplay between spatial and spin anisotropy  is relevant.
In particular, it would be important to explore the crossover from weak
anisotropy
to strong anisotropy of cooperative phenomena
such as those near Lifshitz points.


\begin{thebibliography}{99}
%
\bibitem{kardar}
M. Kardar and R. Golestanian, Rev. Mod. Phys. {\bf71}, 1233 (1999).
%
\bibitem{krech}
M. Krech, {\it The Casimir Effect in Critical Systems} (World Scientific, Singapore, 1994).
%
\bibitem{gambassi}
A. Gambassi, J. Phys. Conf. Ser. {\bf 161}, 012037 (2009).
%
\bibitem{ajdari1991}
A. Ajdari, L. Peliti, and J. Prost,  Phys. Rev. Lett. {\bf 66}, 1481 (1991);
H. Li and  M. Kardar,  Phys. Rev. Lett. {\bf 67}, 3275 (1991);
F. Karimi Pour Haddadan, J. Phys.: Conden. Matter {\bf 29}, 065101 (2017).
%
\bibitem{wil-1}
G.A. Williams, Phys. Rev. Lett. {\bf 92}, 197003 (2004).
%
\bibitem{dohm2011}
V. Dohm, Phys. Rev. E {\bf 84}, 021108 (2011).
%
\bibitem{priv}
V. Privman, A. Aharony, and P.C. Hohenberg, in {\it Phase
Transitions and Critical Phenomena}, edited by C. Domb and J.L.
Lebowitz (Academic, New York, 1991), Vol. 14, p. 1.
%
\bibitem{dohm2018}
V. Dohm, Phys. Rev. E {\bf 97}, 062128 (2018).
%
\bibitem{dohm2008}
V. Dohm, Phys. Rev. E {\bf 77}, 061128 (2008).
%
\bibitem{pri}
V. Privman and M.E. Fisher,  Phys. Rev.  B {\bf  30}, 322 (1984).
%
\bibitem{privman1990}
V. Privman, in {\it Finite Size Scaling and Numerical Simulation of Statistical
Systems}, edited by V. Privman (World Scientific, Singapore,
1990), p. 1.
%
\bibitem{bloete}
H. W. J. Bl\"{o}te, J. L. Cardy, and M. P. Nightingale, Phys. Rev. Lett. {\bf 56}, 742 (1986).
%
\bibitem{affleck}
I. Affleck, Phys. Rev. Lett. {\bf 56}, 746 (1986).
%
\bibitem{dubail}
J. Dubail, R. Santachiara, and T. Emig, EPL {\bf 112}, 66004 (2015); J. Stat. Mech. 033201 (2017).
%
\bibitem{cd2004}
X.S. Chen and V. Dohm, Phys. Rev. E {\bf  70}, 056136 (2004).
%
\bibitem{dohm2006}
V. Dohm, J. Phys.  A {\bf  39}, L 259 (2006).
%
\bibitem{kastening-dohm}
B. Kastening and V. Dohm, Phys. Rev. E {\bf 81}, 061106 (2010).
%
\bibitem{dohm2013}
V. Dohm, Phys. Rev. Lett. {\bf 110}, 107207 (2013).
 %
\bibitem{cardy1987}
J. L. Cardy, in {\it Phase Transitions and Critical Phenomena}, edited by C. Domb and J. L. Lebowitz (Academic, New York, 1987), Vol. 11,  p. 55.
%
\bibitem{cardy1986}
J. L. Cardy, Nucl. Phys. B {\bf 275}, 200 (1986).
 %
\bibitem{cardy1986-270}
J. L. Cardy, Nucl. Phys. B {\bf 270}, 186 (1986).
 %
\bibitem{franc1987}
P. Di Francesco, H. Saleur, and J. B. Zuber, Nucl. Phys. B {\bf 290}, 527 (1987).
%
\bibitem{itz}
For $T\neq T_c$ see C. Itzikson, Nucl. Phys. B (Proc. Suppl.) {\bf 1A}, 185 (1987).
%
\bibitem{cardyconform}
J. L. Cardy, in  {\it Fields, Strings and Critical Pheneomena}, edited by E. Br\'ezin and J. Zinn-Justin (North-Holland Amsterdam, 1990), p.169.
%
\bibitem{franc1997}
P. Di Francesco, P. Mathieu, and D. S\'en\'echal, {\it Conformal Field Theory} (Springer-Verlag New York, 1997).
%
\bibitem{dohm2019}
V. Dohm, Phys. Rev. E {\bf 100}, 050101(R) (2019).
%
\bibitem{dohm2009}
V. Dohm, EPL {\bf 86}, 20001 (2009).
%
\bibitem{footnote}
For $\det {\bf A}\leq0$, the large-distance behavior is affected by the fourth-order moments $\widetilde B_{\alpha \beta \gamma \delta}$ of the couplings $K_{i,j}$, as defined in  Eq. (8.21) of \cite{dohm2008}.
%
\bibitem{chen-zhang}
X.S. Chen and H.Y. Zhang, Int. J. Mod. Phys. B {\bf 21}, 4212 (2007).
%
\bibitem{kim1987}
For an equivalent transformation applied to the eight-vertex and hard hexagon models for the special case $\Omega=\pi/4$ in strip geometry with PBC see D. Kim and P.A. Pearce, J. Phys.  A {\bf  20}, L 451 (1987).
%
\bibitem{suppl}
See the Supplemental Material
for
(i) the derivation of Eqs. (\ref{Casnull}) and (\ref{FcasCFTstrip}) for the critical Casimir force in anisotropic strips,
(ii) the notation of the Jacobi theta functions,
(iii) the analytic expressions for the $d=3$ systems from Ref.~\cite{dohm2018},
(iv) the derivation of the modular invariance of $K_3$.
%
\bibitem{diehl2010}
M. Burgsm\"{u}ller, H.W. Diehl, and M.A. Shpot, J. Stat. Mech. {\bf P}11020 (2010).
%
\bibitem{Indekeu}
J.O. Indekeu, M.P. Nightingale, and W.V. Wang, Phys. Rev. B {\bf 34}, 330 (1986).
%
\bibitem{rud2010}
J. Rudnick, R. Zandi, A. Shackell, and D. Abraham,
Phys. Rev. E {\bf82}, 041118 (2010).
%
\bibitem{FF}
A. E. Ferdinand and M. E. Fisher, Phys. Rev. {\bf 185}, 832 (1969). This case corresponds to $\tau=\rho e^{i\pi/2}=i\rho$.
%
\bibitem{salas}
J. Salas, J. Phys. A {\bf 35}, 1833 (2002).
This case corresponds to $\tau=\rho_{\rm{p}} e^{i\pi/3}$.
%
\bibitem{kastening2012}
B. Kastening,  Phys. Rev. E {\bf 86}, 041105 (2012).
%
\bibitem{izmail}
 N. Sh. Izmailian,  J. Phys. A {\bf 45}, 494009 (2012).
 %
\bibitem{HH2019}
H. Hobrecht and A. Hucht, SciPost Phys. {\bf 7}, 026 (2019).
%
\bibitem{Berker}
R. M. F. Houtappel, Physica {\bf 16}, 425 (1950).
%
\bibitem{Esser}
A. Esser, V. Dohm, and X.S. Chen, Physica  A {\bf  222}, 355
(1995).
%
\bibitem{CDS}
X.S. Chen, V. Dohm, and N. Schultka, Phys. Rev. Lett. {\bf  77}, 3641 (1996).
%
\bibitem{Yurishchev}
M.A. Yurishchev, Phys. Rev. B {\bf 50}, 13533 (1994).
%
\bibitem{Selke2009}
W. Selke and L.N. Shchur,  J. Phys.  A {\bf  38}, L 739 (2005); Phys.Rev. E {\bf 80}, 042104 (2009).
%
\bibitem{lin}
S.-Z. Lin, K. Barros, E. Mun, J.-W. Kim, M. Frontzek, S. Barilo, S. V. Shiryaev,
V. S. Zapf, and C. D. Batista, Phys. Rev.  B {\bf 89}, 220405(R) (2014).

\end{thebibliography}
\end{document}